\documentclass[journal]{IEEEtran}
\IEEEoverridecommandlockouts
\usepackage{cite}
\usepackage{amsmath,amssymb,amsfonts}
\usepackage{algorithmic}
\usepackage{graphicx}
\usepackage{textcomp}
\usepackage{xcolor}
\usepackage{float}
\usepackage{array}
\usepackage{booktabs}
\usepackage{amsmath}
\usepackage{multirow}
\usepackage{subfig}

\usepackage{multirow}     %
\usepackage{booktabs}     %
\usepackage{amsmath,amssymb,amsfonts}
\usepackage{rotating}
\usepackage{adjustbox}
\usepackage{graphicx}

\begin{document}

\bstctlcite{IEEEexample:BSTcontrol}
\title{ACNPU: A 4.75TOPS/W 1080P@30FPS Super Resolution Accelerator with Decoupled Asymmetric Convolution\\
}
\author{\IEEEauthorblockN{Tun-Hao Yang, and Tian-Sheuan Chang, \textit{Senior Member, IEEE}}
\thanks{This work was supported by the National Science and Technology Council, Taiwan, under Grant 111-2622-8-A49-018-SB, 110-2221-E-A49-148-MY3, and 110-2218-E-A49-015-MBK. The authors are with the Institute of Electronics, National Yang Ming Chiao Tung University, Taiwan (e-mail: dhyang.ee09@nycu.edu.tw, tschang@nycu.edu.tw) }%
\thanks{Manuscript received XXXX XX, 2022; revised XXXX XX, XXXX.}
}
\maketitle

\begin{abstract}%
Deep learning-driven superresolution (SR) outperforms traditional techniques but also faces the challenge of high complexity and memory bandwidth. This challenge leads many accelerators to opt for simpler and shallow models like FSRCNN, compromising performance for real-time needs, especially for resource-limited edge devices.
This paper proposes an energy-efficient SR accelerator, ACNPU, to tackle this challenge. The ACNPU enhances image quality by 0.34dB with a 27-layer model, but needs 36\% less complexity than FSRCNN, while maintaining a similar model size, with the \textit{decoupled asymmetric convolution and split-bypass structure}. The hardware-friendly 17K-parameter model enables \textit{holistic model fusion} instead of localized layer fusion to remove external DRAM access of intermediate feature maps. The on-chip memory bandwidth is further reduced with the \textit{input stationary flow} and \textit{parallel-layer execution} to reduce power consumption. Hardware is regular and easy to control to support different layers by \textit{processing elements (PEs) clusters with reconfigurable input and uniform data flow}. The implementation in the 40 nm CMOS process consumes 2333 K gate counts and 198 KB SRAMs. The ACNPU achieves 31.7 FPS and 124.4 FPS for ×2 and ×4 scales Full-HD generation, respectively, which attains 4.75 TOPS/W energy efficiency.

\end{abstract}

\section{Introduction}

Deep learning-based SR models~\cite{dong2015image,lai2017deep, zhang2018image, liu2020residual} are getting popular in recent years due to its superior performance over traditional approaches. As these models evolve, becoming deeper, broader, and more intricate, they also become more computationally demanding and consume more memory bandwidth. The computational load further intensifies with high-definition (HD) or larger input sizes and the corresponding intermediate feature maps. This poses a great challenge for real-time low-power applications on resource-limited edge devices, which demands co-design of hardware acceleration and lightweight model designs.

Several hardware accelerators for SR have been introduced\cite{9223656, 9073972, 9159619, huang2019ecnn, 9401206, BSRA} tailored for real-time HD applications. Yet, many current solutions prioritize ease of hardware design over image quality by adopting simpler network architectures like the commonly employed FSRCNN or FSRCNN-s\cite{dong2016accelerating} and their basic variants. This choice, driven by complexity concerns, often results in the compromised quality of the reconstructed images. Moreover, the significant challenge posed by memory bandwidth remains unaddressed in many designs. The prevalent layer-by-layer processing technique often necessitates storing intermediate data in DRAM and subsequently reloading them for each layer or demands a substantial buffer for temporary storage. Both approaches are impractical given the vast feature maps. To address the bandwidth dilemma, works like ~\cite{huang2019ecnn, 9159619} have employed tile-based layer fusion. Although this reduces bandwidth requirements, substantial intermediate data bandwidth and buffer storage are still needed, especially for tile boundary data, owing to their model configurations.

\begin{figure}[tbhp]
    \centering
    \includegraphics[height=!,width=0.8\linewidth,keepaspectratio=true]{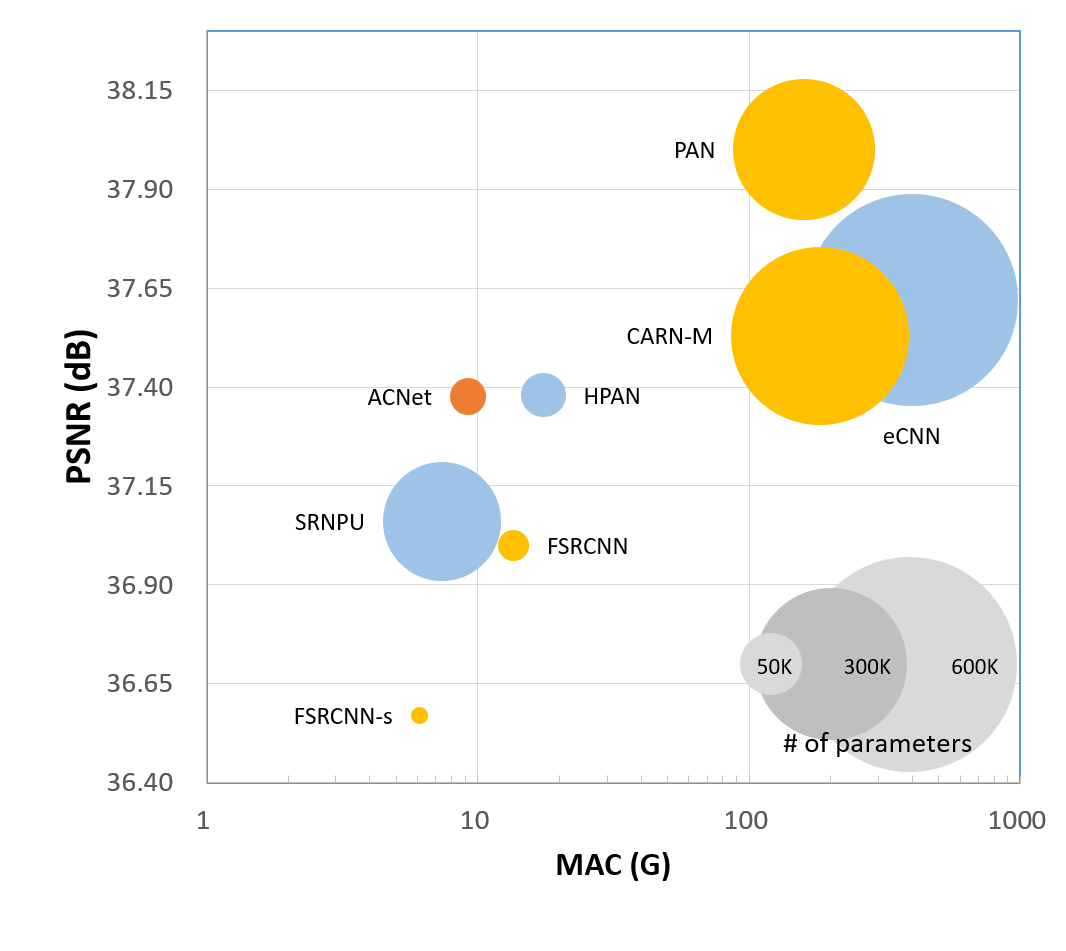}
    \caption {Comparison of different SR models in terms of model size, multiply-accumulate (MAC) and peak signal-to-noise ratio (PSNR). The yellow dots are software works and the blue dots are hardware models. The orange dot is our ACNet model.}
    \label{acnet compare}
\end{figure}

Several lightweight SR models have been developed to balance image quality against computational complexity. Fig.\ref{acnet compare} contrasts the operations, parameters, and performance of these models. Within this comparison, FSRCNN-s\cite{dong2016accelerating} offers compactness but at the cost of image quality. SRNPU~\cite{9159619}, with over 100K parameters, employs dynamic processing, segmenting image tiles to manage processing workload effectively. HPAN~\cite{BSRA} introduces a simplified pixel attention model and non-overlapping block processing for $\times2$ scaling. However, it wrestles with computational demands and artifacts at $\times4$ scaling. Additionally, SR models incorporate lightweight techniques such as depthwise convolution\cite{8429522}, group convolution~\cite{DBLP:journals/corr/abs-2105-10288}, and asymmetric convolution\cite{ding2019acnet, tian2021asymmetric}. These models, designed primarily for software, possess intricate architectures that aren't friendly for hardware implementation, leading to increased buffer demands. Prior models also tend to intertwine asymmetric convolutions or combine them with standard ones, which hinders optimizing the benefits of asymmetric convolution and raises buffer requirements.

Addressing the above issues, this paper introduces a real-time SR neural processing unit, ACNPU, optimized through a synergy of hardware and software. The SR accelerator enhances image quality by 0.34dB using a 27-layer architecture, ACNet, yet it boasts 36\% less complexity than the FSRCNN, while maintaining a similar model size. This is achieved using a \textit{decoupled asymmetric convolution and split-bypass structure}. Designed to be hardware-friendly, the model leverages regular structures and localized connections, avoiding long connections to save buffer cost. Its 17K model size allows the accelerator to bypass external memory access for intermediate feature maps, thanks to the \textit{holistic model fusion}. Additional efficiencies are realized by parallel execution of 1x3/1x1 layers and a \textit{local input stationary flow}, reducing internal memory access. The core of this system is built on regular processing element (PE) clusters, which have a uniform data flow and reconfigurable input across layers. The implementation is capable of real-time full-HD processing, offering an energy efficiency of 4.75 TOPS/W.

The remainder of the paper is organized as follows. Section II presents the proposed ACNet model. Section III shows the hardware design. Section IV illustrates the experimental results. Finally, we conclude this paper in Section V.

\section{Proposed ACNet for superresolution}

\subsection{Proposed ACNet}

The primary objective of the target model is to enhance image quality using hardware-optimized structures. It aims to achieve this while maintaining a model size and complexity comparable to other existing SR accelerators. Notably, previous lightweight SR models, optimized primarily for software, still demand parameters nearing the mega scale. In contrast, the models typically employed in other SR accelerators don't emphasize quality optimization.

\begin{figure}[htbp]
    \centering
    \includegraphics[height=!,width=0.8\linewidth,keepaspectratio=true]{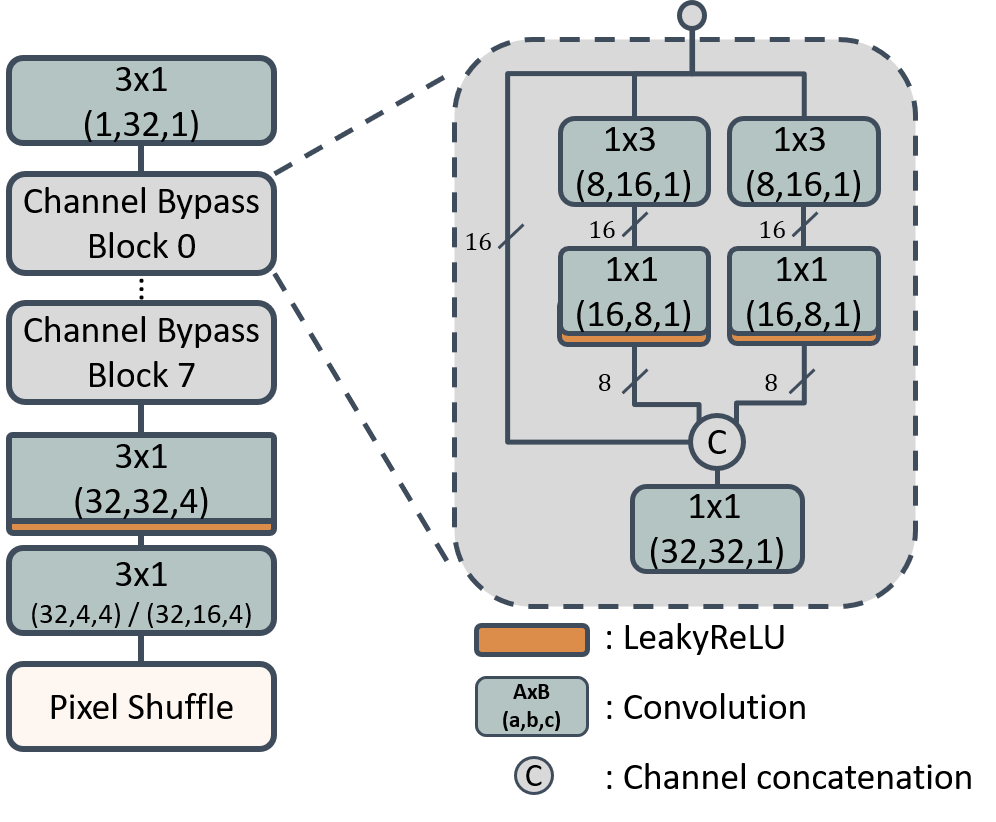}
    \caption {The proposed ACNet. The convolution notation, A×B, (a, b, c), is kernel size A×B, input channel a, output channel b, and group number c, respectively.}
    \label{ACNet model}
\end{figure}

Fig.~\ref{ACNet model} shows the proposed model, ACNet, to meet the above goal, which contains one 3x1 convolution layer, eight channel-bypass blocks (CBBs) with 1x3/1x1/1x1 layers, two 3x1 convolution layers and one pixel shuffle layer for final reconstruction output. 
For quality, this model cascades eight CBBs to extract features and form a deeper model than before for better performance. However, even with deeper models, we reduce the complexity and model size with \textit{decoupled asymmetric convolution and split-bypass structure}, which is also hardware-friendly. The model uses local short-cut connections instead of global short-cut connections to avoid large buffers. 

The asymmetric convolution uses kernel sizes of 1xK or Kx1 instead of KxK to reduce complexity. However, the adopted asymmetric convolution, unlike the previous tightly coupled 1xK/Kx1 arrangement, is decoupled to two ends of the model. The top and bottom layers use vertical directions (3x1), while the central layers (CBB) use horizontal directions (1x3). CBBs are our main backbone for feature extraction. However, if we use only one type of asymmetric convolution, the model cannot learn the information from the other dimension. Thus, we use vertical convolution in the first and last two stages to achieve better performance without a large buffer size.

The CBB consists of a split-bypass operation that uses 1×3 and 1×1 convolutions and bypasses channels as a balance between complexity and performance. In these CBBs, we only use one direction of asymmetric convolution (1×3) instead of two directions as in other models to save boundary buffer size and ease hardware design. In CBBs, the input channels are divided into three groups with 16 channels in each group, where two groups will be processed and one group is bypassed to save computation while preserving performance. At the end of the block, a 1×1 convolution is applied to merge the features of all different channels for better reconstruction.

\subsection{Analysis of ACNet}
\begin{figure}[htbp]
    \centering
    \includegraphics[height=!,width=0.8\linewidth,keepaspectratio=true]{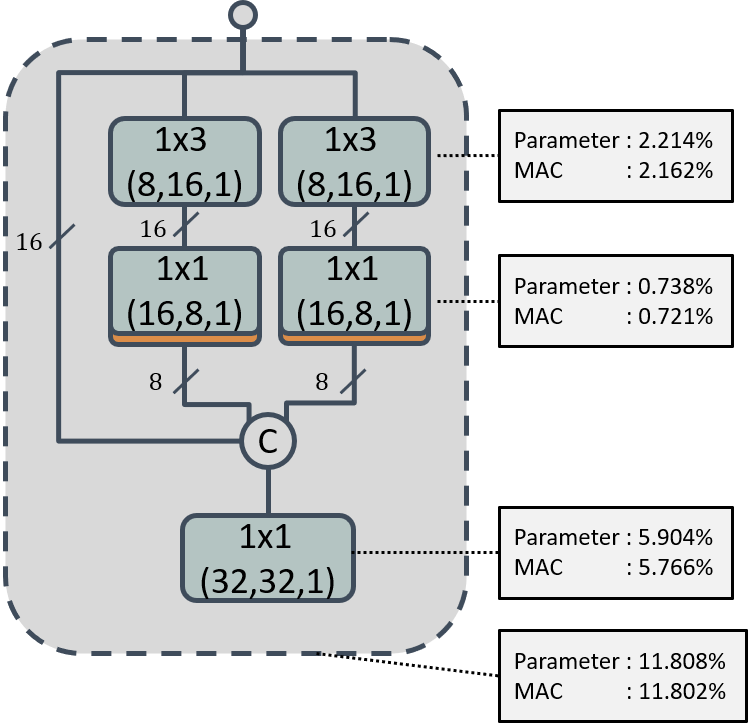}
    \caption {Parameters and operations detail of the CBB.}
    \label{parameter operation CBB}
\end{figure}

\begin{figure}[htbp]
    \centering
    \includegraphics[height=!,width=0.8\linewidth,keepaspectratio=true]{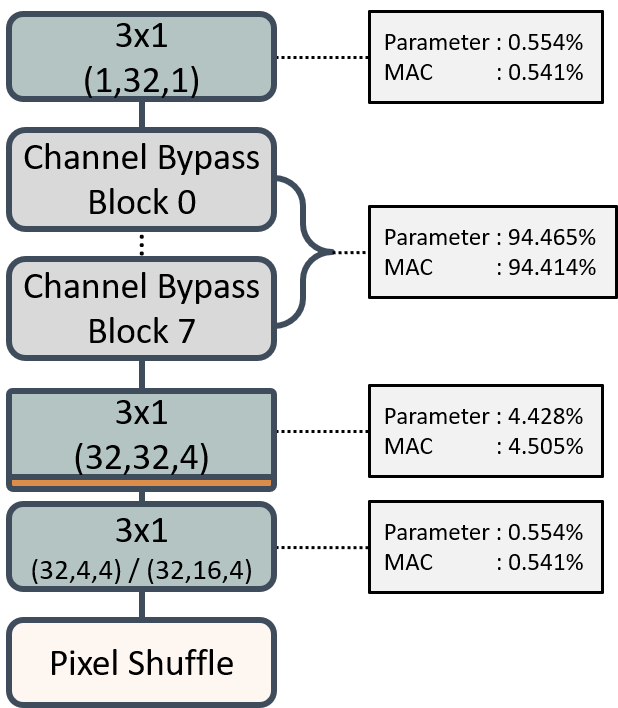}
    \caption {Parameters and operations of the overall model.}
    \label{parameter operation ACNet}
\end{figure}

Fig.~\ref{parameter operation CBB} and Fig.~\ref{parameter operation ACNet} show details of the basic block and the whole model. In Fig.~\ref{parameter operation CBB}, due to channel bypass, the operation and parameter are reduced by 34 \% compared to the one without channel pass. The reason for using a normal 1×1 convolution at the end of the block is to improve the performance of the reconstruction. The last two 3×1 convolution layers adopt group convolution to reduce about 75 \% of parameters and operations. In addition, asymmetric convolution can reduce 44 \% of operations and 45 \% of parameters. Fig.~\ref{parameter operation ACNet} shows the parameters and operations of the overall model, where the CBBs occupy 94\%. 

Fig.~\ref{acnet compare} shows the ball chart of parameters and operation count versus performance in the Set5 test data set. The proposed ACNet achieves a performance comparable to HPAN\cite{BSRA}, but reduces the parameters and operation to 67\% and 53\% of HPAN. Compared to \cite{8429522}, which also uses asymmetric convolution, our performance is 0.68 dB higher.

\subsection{Quantization format}
Most previous work uses the fixed point (FXP) format in their architectures for a lower cost, but the floating-point (FP) format is better for performance. For example, in \cite{9073972} and HPAN\cite{BSRA}, their FXP performance is decreased by 0.3 and 0.2 dB, respectively, compared to their FP counterpart. The situation results from the quantization error and the outlier values. \cite{9401206} uses both formats in its design to handle some outliers and decrease external memory access. Instead of using hybrid processing, all features and weights in our design are quantized to 13 bits (S1E5M7) and 10 bits (S1E5M4) in the FP format, respectively, as a trade-off between performance and cost, where S: sign, E: exponent, M: mantissa. 

Table~\ref{The results of two quantization formats} shows the quantization in FP and FXP format with different bit lengths. We can see that the results of the two methods in the scaling factor $\times$2 are similar, but FP13 is better in the scaling factor $\times$4 due to a wider range of values than the other to handle more outliers that can damage the performance of the model.

\begin{table}[htbp]
\centering
\caption{The results of two quantization formats.}
\label{The results of two quantization formats}
\begin{tabular}{|ccc|}
\hline
\multicolumn{1}{|c|}{Q-Type} & Set5  & B100  \\ \hline
\multicolumn{3}{|c|}{x2}                     \\ \hline
\multicolumn{1}{|c|}{FP32}   & 37.41 & 31.73 \\ 
\multicolumn{1}{|c|}{FXP13}  & 37.35 & 31.71 \\
\multicolumn{1}{|c|}{FP13}   & 37.37 & 31.71 \\ \hline
\multicolumn{3}{|c|}{x4}                     \\ \hline
\multicolumn{1}{|c|}{FP32}   & 31.22 & 27.20 \\
\multicolumn{1}{|c|}{FXP13}  & 30.94 & 27.11 \\
\multicolumn{1}{|c|}{FP13}   & 31.20 & 27.20 \\ \hline
\end{tabular}

\end{table}

\section{System Architecture of ACNPU}
\subsection{Design Challenges and Proposed Solutions}
Due to the large input size and deep SR models, traditional SR accelerators face three design challenges: hardware cost, external memory access, and internal memory access.  In this work, we markedly reduce the hardware cost associated with MAC operations through our proposed lightweight, hardware-optimized ACNet. The central issue then becomes to execute the necessary 3x1 / 1x3 / 1x1 layers cost-effectively, which we address using \textit{PE clusters with reconfigurable input and uniform data flow}. 

Regarding external memory access, the compact nature of our model, sized at 17K, allows us to eliminate external memory access of intermediate feature maps. This is achieved through a \textit{holistic model fusion} strategy rather than a localized layer fusion as discussed in~\cite{7783725}. This fusion processes a non-overlapping 3$\times$192 tile of the image through the entire model before proceeding to the subsequent tile. This continues until the full super-resolution image is synthesized. During the process, the small model is easily stored on the chip to save frequent access.  All boundary partial sums of tiles are stored in boundary SRAMs, operating in a ping-pong buffer fashion. This storage method helps reconstruct the image without manifesting checkerboard patterns and reduces additional DRAM access as compared to the approach in \cite{9159619}.

To reduce internal memory access, our design executes  1$\times$3 and 1$\times$1 layers in the CBB in parallel, facilitating direct data transfers between PEs without the need for intermediary storage. Additionally, the design employs a \textit{local input stationary flow} that temporarily stores data from local SRAMs, eliminating the need for constant accesses over several computing cycles.

\begin{figure}[htbp]
    \centering
    \includegraphics[height=!,width=1.0\linewidth,keepaspectratio=true]{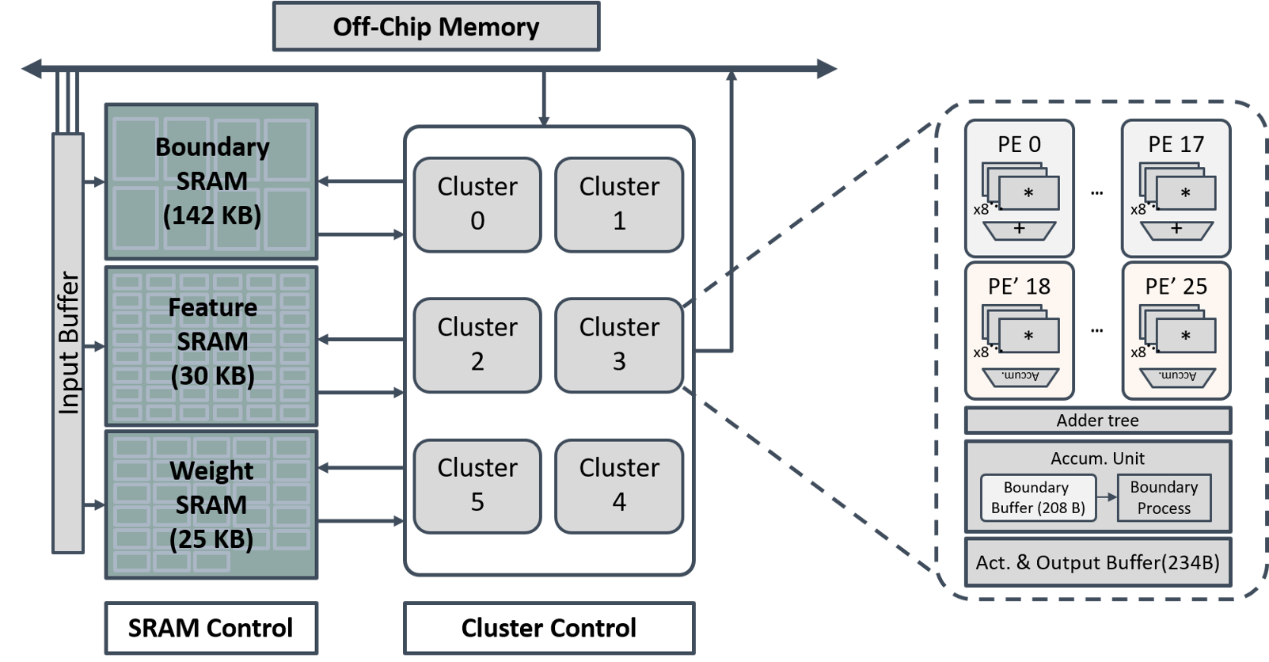}
    \caption {The proposed system architecture.}
    \label{ACNPU_overall}
\end{figure}

Fig.~\ref{ACNPU_overall} shows the overall architecture that includes six clusters for computing and three SRAM buffers for storage. Depending on the requirement, this design can handle full HD upscaling for $\times$2 and $\times$4 and any number of CBBs within 8. The following will introduce details of this design.

\subsection{Detailed Architecture of the cluster}
\begin{figure}[htbp]
    \centering
    \includegraphics[height=!,width=1.0\linewidth,keepaspectratio=true]{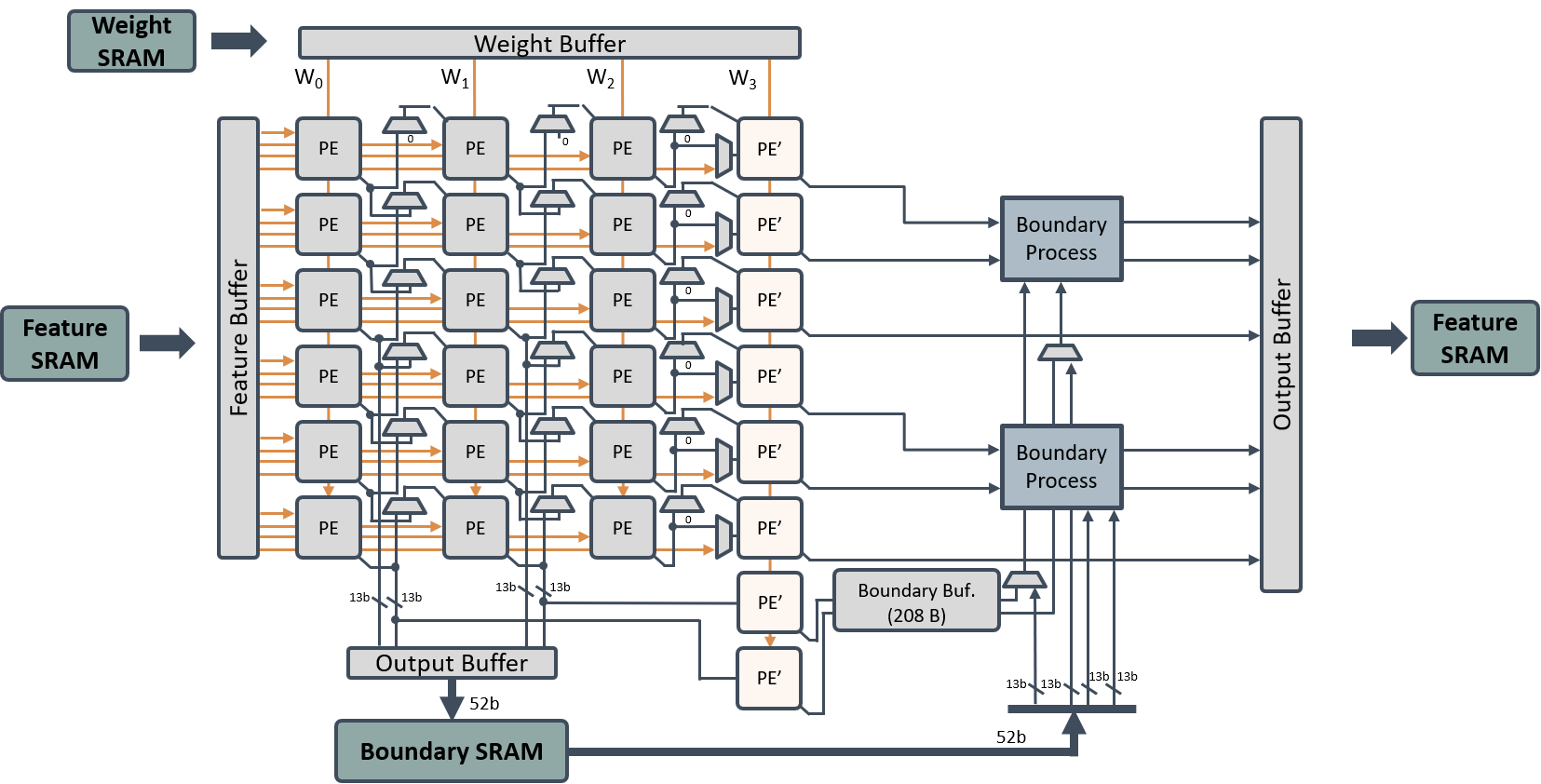}
    \caption {Detail of the processing cluster.}
    \label{cluster_overall}
\end{figure}

Fig.~\ref{cluster_overall} illustrates the cluster's architecture. This architecture comprises 18 \textit{PE} and 8 \textit{PE'} modules. These modules enable the support of all convolution types present in the model through input reconfiguration. The cluster employs the \textit{local input stationary flow}. This flow buffers the feature maps across several computing cycles to save recurrent access. Input and weights are broadcast to each MAC for simultaneous processing. Each module then accumulates the results. The partial sum results located at the boundary are combined at the two boundary processing blocks. This combination uses previously stored data either from the boundary SRAMs or registers, yielding the final output. The following sections provide a detailed description of the data flow.

\begin{figure}[htbp]
\subfloat[PE]{\includegraphics[height=!,width=0.45\linewidth,keepaspectratio=true]{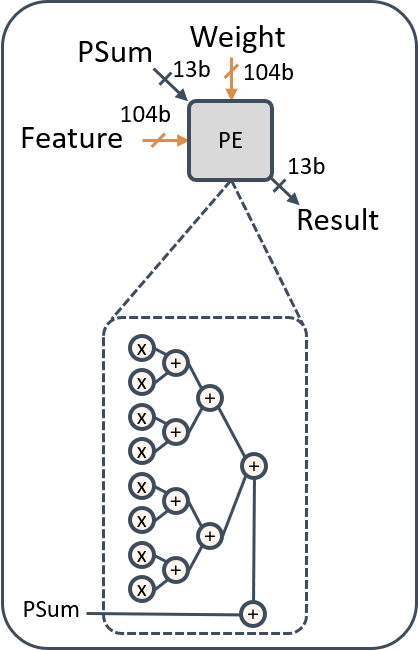}}
\subfloat[PE']{\includegraphics[height=!,width=0.45\linewidth,keepaspectratio=true]{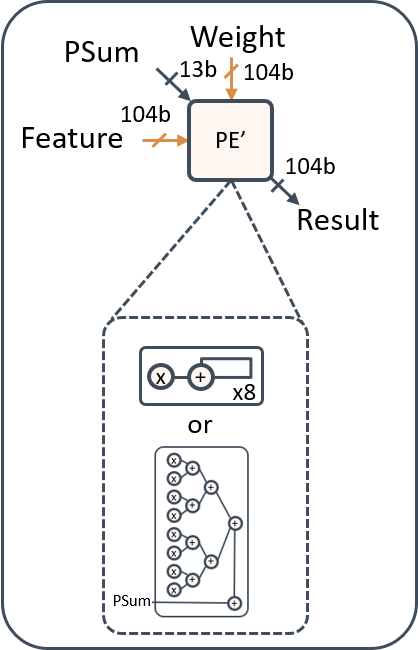}}

\caption {Details of the \textit{PE} and the \textit{PE'}.}
\label{Details of the PE and the PE'.}
\end{figure}

For the two types of processing elements, \textit{PE} has only one mode, and \textit{PE'} has two modes to support different layers. For \textit{PE} as shown in Fig.~\ref{Details of the PE and the PE'.} (a), eight inputs and weights will be multiplied, and then the eight results and partial sums of the other \textit{PEs} will be accumulated together to generate the results. \textit{PE'} as shown in Fig.~\ref{Details of the PE and the PE'.} (b) is similar to \textit{PE} but could also accumulate values from different cycles. 

The following shows the four operating modes in PE clusters to support different layers with reconfigurable input and uniform data flow for easy control and low area overhead.

\subsubsection{Mode 1: 3×1 convolution}

\begin{figure*}[htbp]
\subfloat[Dataflow of the cluster]{\includegraphics[height=!,width=0.47\linewidth,keepaspectratio=true]{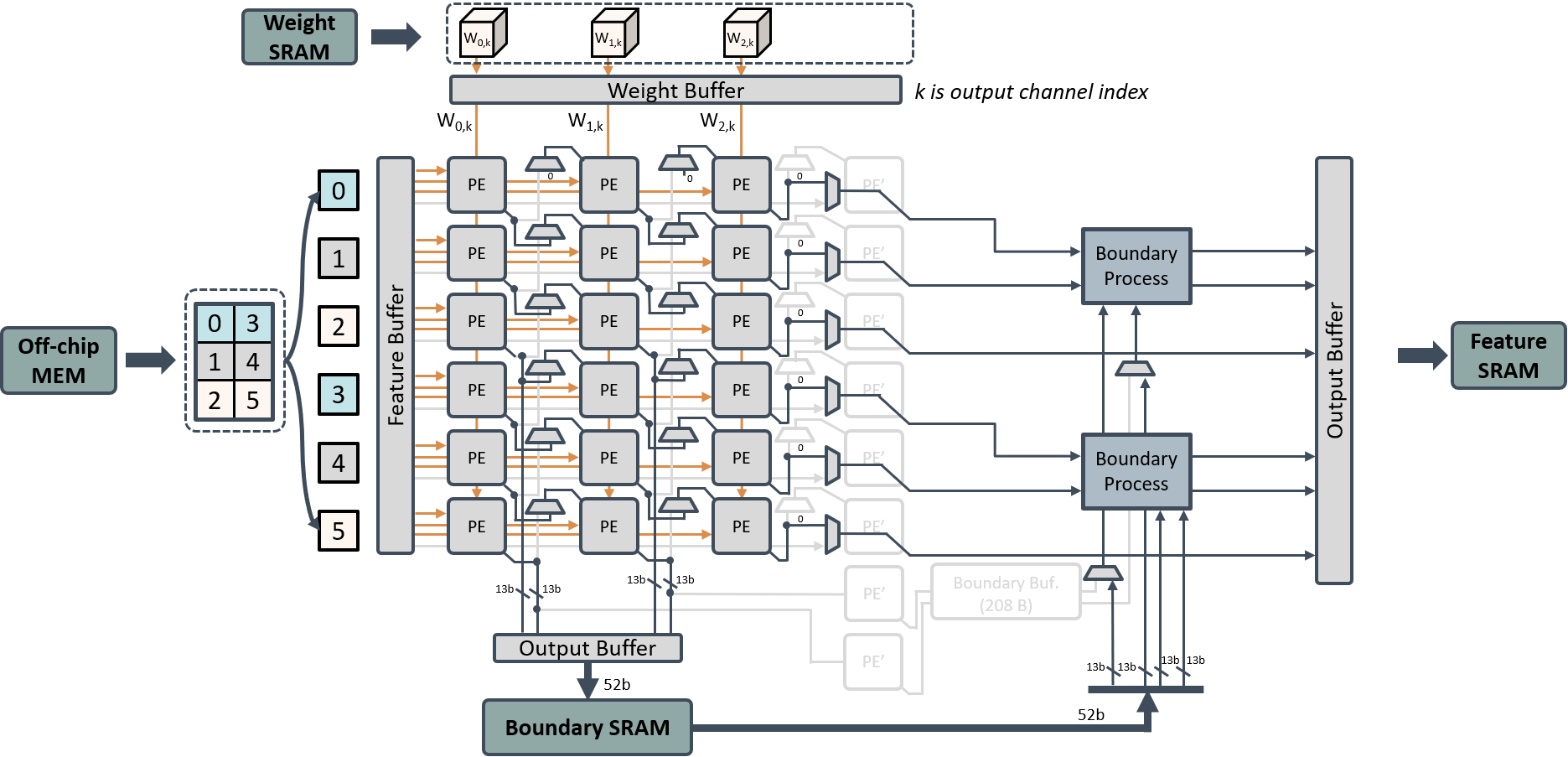}}
\hfill
\subfloat[Scheduling of the input data]{\includegraphics[height=!,width=0.47\linewidth,keepaspectratio=true]{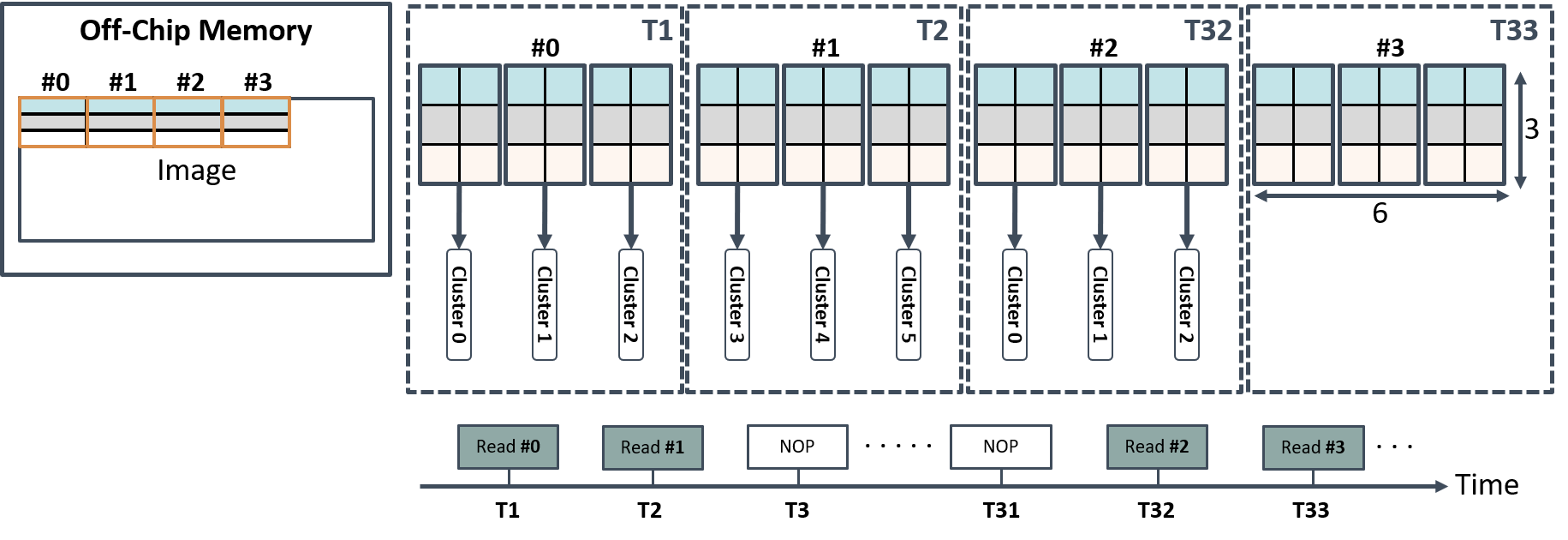}}

\caption {The dataflow of the 3$\times$1 convolution layer.}
\label{The dataflow of first 3x1 convolution}
\end{figure*}

Fig.\ref{The dataflow of first 3x1 convolution} shows the data flow and access to the input data for the vertical (3×1) convolution in the initial layer. As illustrated in Fig.\ref{The dataflow of first 3x1 convolution} (b), during the first cycle \emph{T1}, the primary 3×6 input pixels, labeled as \emph{\#0}, are fetched from the off-chip memory. These pixels are then stored within the input buffers and sent to the cluster 0 through 2. Subsequently, in the second cycle, the second input \emph{\#1} is retrieved and reserved for the cluster 3 through 5. To minimize the input buffer access frequency, these inputs remain constant and undergo updates every 32 cycles, exemplifying our input stationary flow. Meanwhile, as shown in Fig.~\ref{The dataflow of first 3x1 convolution} (a), the weight buffer is refreshed in each cycle using data from the weight SRAM. The weight retrieval sequence is cyclical, beginning with output channel 0 and ending at output channel 31, continuing until the entire input tile undergoes processing.

Within the flow described, Fig.~\ref{The dataflow of first 3x1 convolution} (a) highlights that six inputs for a single cluster are horizontally broadcast to 18 \textit{PEs}. These are then multiplied by vertically broadcast weights. The resulting products of this multiplication are diagonally cumulated. In particular, while each PE accommodates eight inputs, only one from the input image is available. This limitation reduces \textit{PE's} utilization to a mere 8.6\%, representing a significant under-utilization. However, the execution time of this mode is relatively short compared to that of the other three modes, as shown later.

\subsubsection{Mode 2: parallel execution of 1$\times$3 and 1$\times$1 convolutions}

\begin{figure*}[htbp]
\subfloat[Dataflow of the cluster]{\includegraphics[height=!,width=0.47\linewidth,keepaspectratio=true]{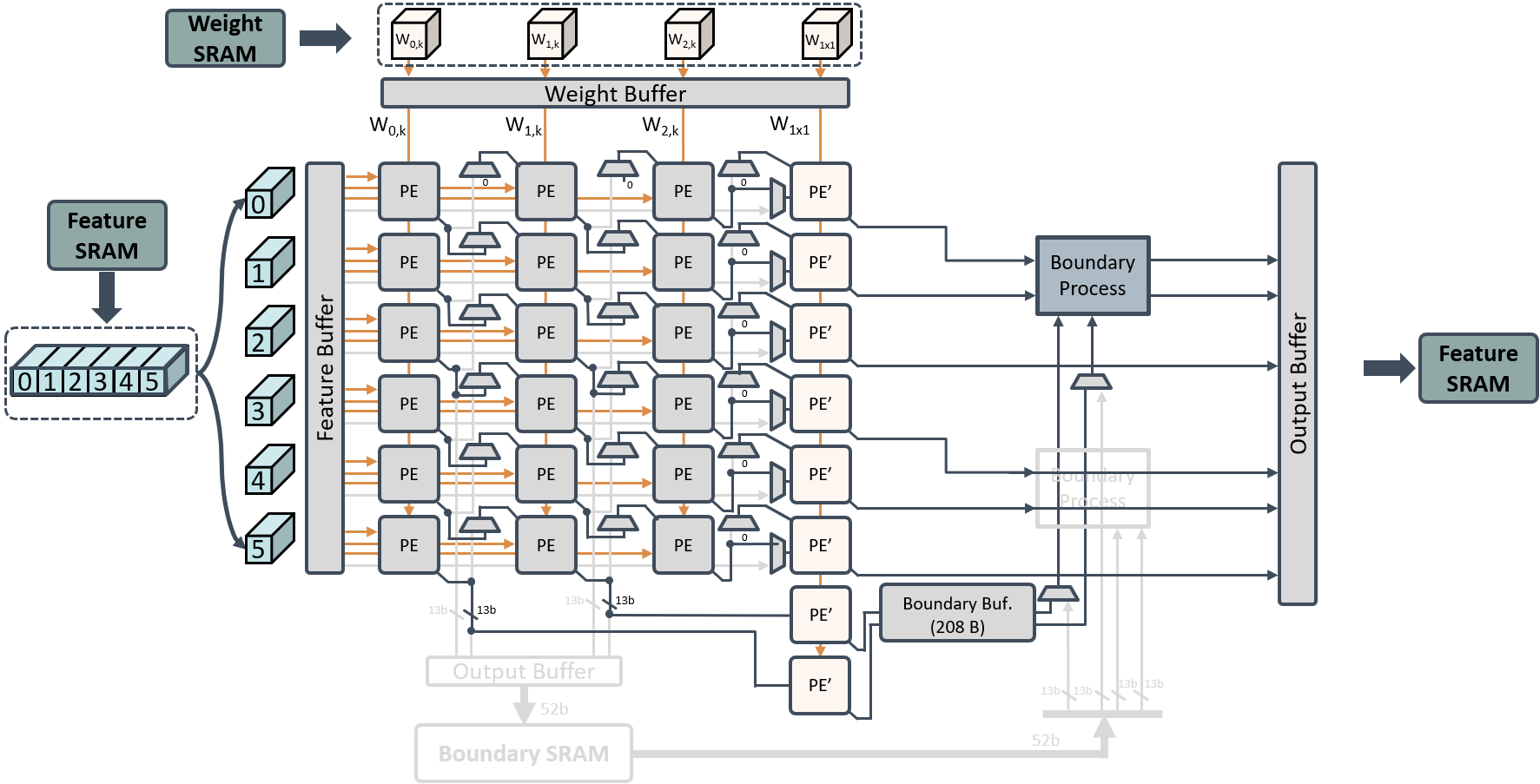}}
\hfill
\subfloat[Scheduling of the input data]{\includegraphics[height=!,width=0.47\linewidth,keepaspectratio=true]{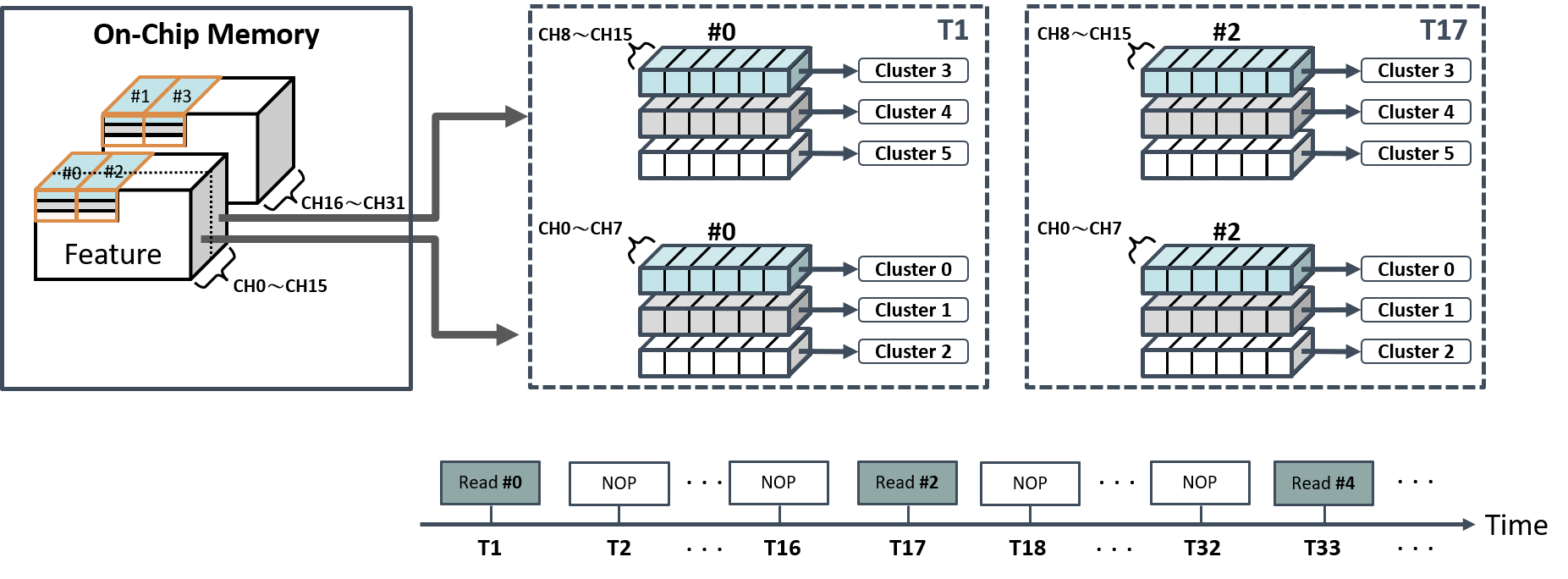}}

\caption {The data flow of the 1$\times$3 and 1$\times$1 convolution layers in CBB.}
\label{The data flow of 1x3 and 1x1 convolution in CBB}
\end{figure*}

Fig.\ref{The data flow of 1x3 and 1x1 convolution in CBB} illustrates the data flow and input data access mechanisms for the 1$\times$3 and 1$\times$1 convolutions within the CBB block. Given that the 1×3 and 1×1 convolutions together account for 46\% of the computational complexity in this model, they are cascaded—1×3 convolution at 18 \textit{PEs} and 1×1 convolution at 8 \textit{PE's}—to curtail internal memory access, leading to a reduction in power consumption. As highlighted in Fig.\ref{The data flow of 1x3 and 1x1 convolution in CBB} (b), the 6x3x16 input features are fetched from on-chip memory every 16 cycles: channel 0$\sim$7 are designated for the cluster 0$\sim$2, while channel 8$\sim$15 are reserved for the cluster 3$\sim$5. Weight updates in this mode occur cyclically, analogous to mode 1. Both \textit{PE} and \textit{PE'} modules are actively utilized here. While the weight sequencing in \textit{PEs} (for the 1×3 convolution) mirrors that of mode 1, the \textit{PE's} (for the 1×1 convolution) rotate their weight order from channel 0 to 15.

Following the stipulated access sequence, weights are broadcast vertically to the \textit{PEs} (for 1×3 convolutions) and the \textit{PE's }(for 1×1 convolutions). Inputs are broadcast horizontally across 18 \textit{PEs} for the 1×3 convolution, with the resultant outputs relayed to \textit{PE's} for the subsequent 1×1 convolution. As illustrated in Fig.~\ref{The data flow of 1x3 and 1x1 convolution in CBB} (a), these partial sums spanning different channels are then aggregated. Finally, the partial boundary sums derived from the bottom two \textit{PE's} are retained in the boundary buffer, poised to be amalgamated with other boundary values.

\subsubsection{Mode 3: 1×1 fusion convolution layer}
\begin{figure*}[htbp]
\subfloat[Dataflow of the cluster]{\includegraphics[height=!,width=0.47\linewidth,keepaspectratio=true]{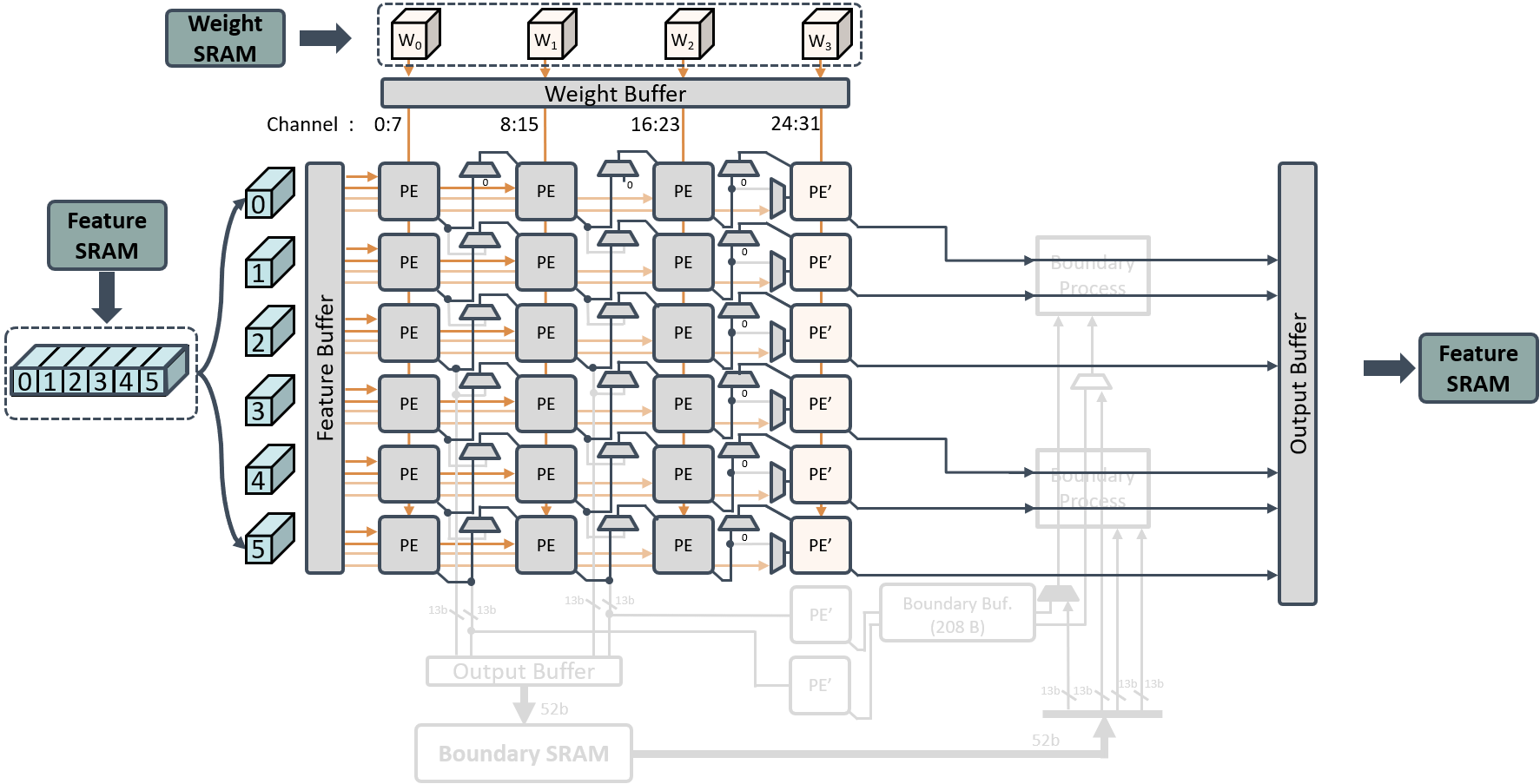}}
\hfill
\subfloat[Scheduling of the input data]{\includegraphics[height=!,width=0.47\linewidth,keepaspectratio=true]{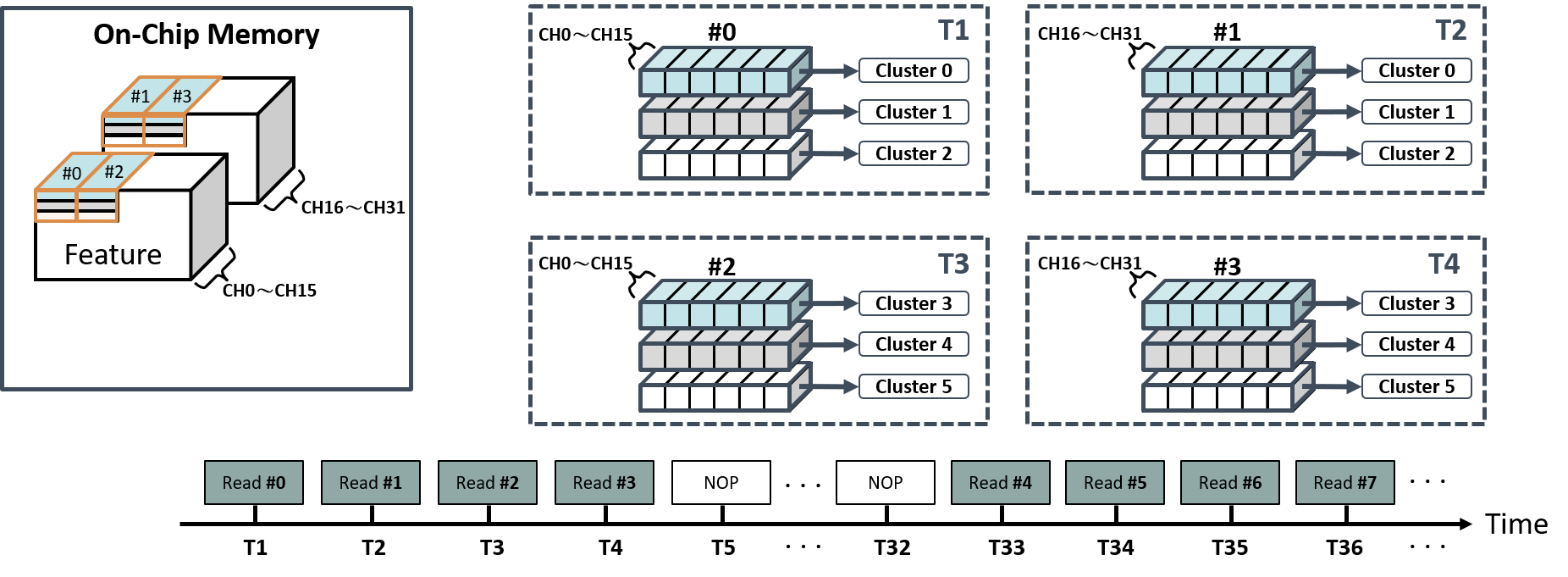}}

\caption {The data flow of the 1$\times$1 fusion convolution layers in CBB.}
\label{The data flow of 1x1 fusion convolution in CBB}
\end{figure*}

As for 32 channels of the 1×1 convolution layer in the CBB, Fig.~\ref{The data flow of 1x1 fusion convolution in CBB} shows the data flow and the access of the input data. As in Fig.~\ref{The data flow of 1x1 fusion convolution in CBB} (b), the 3x6x16 features are read from on-chip memory and are updated every 32 cycles. The data \emph{\#0} and \emph{\#1} in the channel 0 to 15 and 16 to 31 are distributed to the cluster 0$\sim$2  at the \emph{T1} and \emph{T2} cycles, respectively. Similar operations are applied to cluster 3$\sim$5 in the next two cycles for the data \emph{\#2} and \emph{\#3}.  

The data flow is as follows. In this mode, \textit{PE} and\textit{ PE'} have the same operation. To increase hardware utilization, unlike other modes, the first 16 channels are given to the 12 left \textit{PEs}, and the last 16 channels are passed to the 6 right \textit{PEs} and 6 \textit{PE's}. This is the reason why we need independent input feature buffers for every PE instead of row-wise broadcasting. Although the hardware cost is increased, we can increase utilization by double. The order of weights is read circularly from output channel 0 to output channel 31, which are broadcast to the \textit{PEs} and \textit{PE's}.

\subsubsection{Mode 4: 3×1 group convolution}
This mode is similar to mode 1, except that the input is updated every 8 cycles
due to its group operations. If the output channel is $g=32$, the cycle count to update the input is given by $g/4 = 8$ in our case.

\subsection{Boundary SRAM}

\begin{table}[htbp]
\centering
\caption {Analysis of three vertical convolutions in the ACNet.}
\label{The SRAM and Bandwidth}
\begin{tabular}{|c|c|c|}
\hline
\multicolumn{1}{|l|}{} & SRAM (KB) & Off-Chip Bandwidth (MB/s) \\ \hline
CONV\_3x1\_1           & 57        & 1.68                      \\
CONV\_3x1\_2           & 57        & 1.68                      \\
CONV\_3x1\_3           & 28        & 0.84                      \\ \hline
Total                  & 142       & 4.20                      \\ \hline
\end{tabular}

\end{table}

Table~\ref{The SRAM and Bandwidth} shows the boundary SRAM used in the three 3×1 convolution layers as in Fig.~\ref{ACNet model} by assuming ×2 scaling and FullHD@30fps output. The total buffer size is larger than the feature and weight SRAM due to the ping-pong buffer strategy to keep the hardware busy. A tradeoff to reduce buffer size is to store them in the external memory with the extra DRAM access, which requires an extra 4.2MB/s bandwidth without this boundary buffer.

\section{Experimental Result}
\subsection{Experimental setup}
For the data sets, we use two training datasets, DIV2K and Flickr2K, and five commonly used benchmark datasets, Set5\cite{bevilacqua2012low}, Set14\cite{yang2010image}, B100\cite{martin2001database}, and Urban100\cite{huang2015single}
with PSNR as the evaluation metric.

For model training, we randomly crop 256x256 patches that are from training datasets and apply data augmentation to these patches, such as rotation and flip. The batch size is 128. The L1 loss function is used with the AdamW optimizer for training. We use a step learning scheduler with gamma set as 0.5 for learning rate decay every 200 epochs. The initial learning rate is 5e-3 and the total epochs are 2000. The proposed model is implemented in the PyTorch framework with the NVIDIA DGX Station A100.

\subsection{Model evaluations}
Table ~\ref{Comparison of several light weight models.}  and Table~\ref{Comparison of several light weight models. x4}  show the experimental results for the $\times$2 and $\times$4 scaling. The MAC is calculated assuming that the output image is 1920$\times$1080. These tables select lightweight models for comparison, which can be divided into lightweight models (model parameters $>$ 100K) and ultra-lightweight ones (model parameters $\leq$ 100K).

Compared to the best lightweight model in the table, our performance in B100 is only 0.75dB lower. However, its parameters and operations are 31$\times$ and 25$\times$ different from ours, respectively. Among lightweight models, LWN~\cite{LWN} has a performance similar to ours in Set5 and Set14, but better than ours in B100 and Urban100. The reason is that the images in Urban100 usually have a regular texture, which cannot be handled well because of the less vertical direction information in our model. However, if we take into account the computation cost and simplicity, our simple and powerful model still has its advantages.

\begin{figure}[htbp]
    \centering
    \includegraphics[height=!,width=1.0\linewidth,keepaspectratio=true]{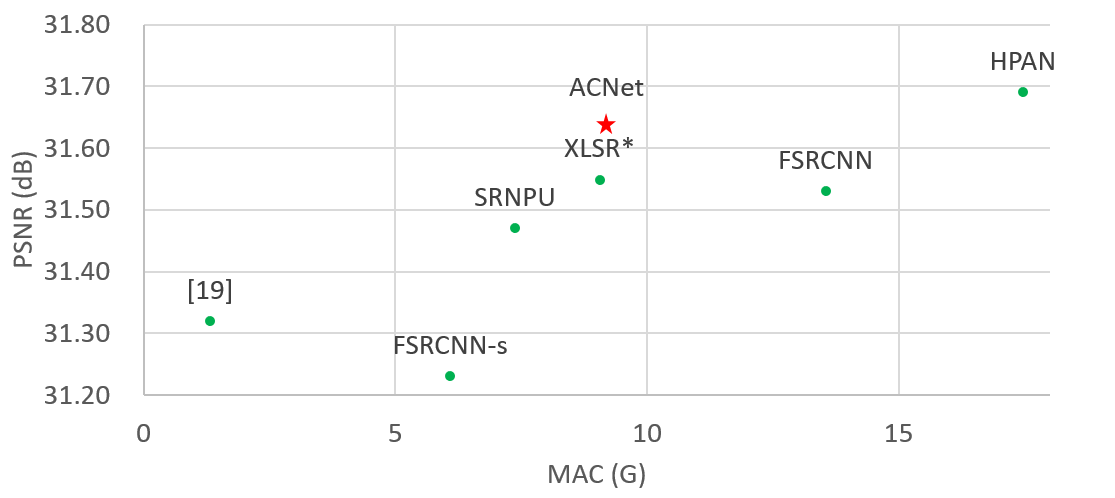}
    \caption {PSNR vs MAC numbers under 20G.}
    \label{Distribution of models whose MAC are below 20G}
\end{figure}

Compared to the ultra-lightweight models, Fig.~\ref{Distribution of models whose MAC are below 20G} shows their performance in B100 and the number of MACs. The red star symbol represents our method. Compared to HPAN, this work shows a slight loss in PSNR, but reduces almost 50 \% of MAC operations.  For XLSR, because the original model in \cite{DBLP:journals/corr/abs-2105-10288} is designed for scaling in $\times$2, we modified the number of output channels in the upsampling stage and the modified model is marked as XLSR*.  The parameter of XLSR* is the same as ours, but our model has better or equivalent performance in all test datasets. Compared to \cite{8429522} with the coupled asymmetric convolution, our model has better results. SRNPU uses two FSRCNN-like models, and thus has comparable performance to FSRCNN but with reduced operation count. However, its parameter is 10$\times$ larger than ours and its performance is 0.3 dB lower than FSRCNN for the upscaling factors $\times$4 in Set5.

\begin{table*}[htbp]
\begin{minipage}{\columnwidth}
\centering
\caption{The quantitative results of several light weight methods($\times$2). The bold one represents our method.}
\label{Comparison of several light weight models.}
\begin{tabular}{|c|c|c|c|c|c|c|c|}
\hline
\textbf{Model} & \textbf{Size(K)} & \textbf{GMAC} & \textbf{Set5}  & \textbf{Set14} & \textbf{B100}  & \textbf{U.100}      \\ \hline
Bicubic        & -               & -                & 33.66          & 30.24          & 29.56          & 26.88          \\ \hline
$^a$M.IDN          & 13              & -                & 37.23          & 32.89          & 29.58          & 30.43      \\ \hline
FSRCNN-s       & 4               & 6                & 36.57          & 32.28          & 31.23          & -              \\ \hline
{[}19{]}       & 3               & 1                & 36.66          & 32.52          & 31.32          & 29.34          \\ \hline
SRCNN          & 57              & 119              & 36.66          & 32.45          & 31.36          & 29.50          \\ \hline
$^a$SRNPU          & 183             & 7                & 37.06          & 32.62          & 31.47          & -          \\ \hline
FSRCNN         & 13              & 14               & 37.00          & 32.63          & 31.53          & 29.88          \\ \hline
XLSR$^*$          & 17              & 9                & 37.17          & 32.78          & 31.55          & 29.89       \\ \hline
\textbf{ACNet} & \textbf{17}     & \textbf{9}       & \textbf{37.34} & \textbf{32.78} & \textbf{31.64} & \textbf{30.21} \\ \hline
$^a$HPAN           & 26              & 17               & 37.38          & 32.91          & 31.69          & 30.29      \\ \hline
LapSRN         & 813             & 67               & 37.52          & 33.08          & 31.80          & 30.41          \\ \hline
LWN            & 286             & 104              & 37.38          & 32.93          & 31.85          & 31.04          \\ \hline
VDSR           & 665             & 1,225            & 37.53          & 33.03          & 31.90          & 30.76          \\ \hline
CARN-M         & 412             & 182              & 37.53          & 33.26          & 31.92          & 31.23          \\ \hline
$^a$eCNN           & 583             & 398              & 37.62          & 33.17          & 31.93          & 30.60      \\ \hline
DRRN           & 297             & 13,594           & 37.74          & 33.23          & 32.05          & 31.23          \\ \hline
MADNet         & 878             & 187              & 37.85          & 33.39          & 32.05          & 31.59          \\ \hline
IDN            & 579             & 280              & 37.83          & 33.30          & 32.08          & 31.27          \\ \hline
T-FMB          & 649             & -                & 37.90          & 33.45          & 32.11          & 31.83          \\ \hline
RFDN           & 534             & 284              & 38.05          & 33.68          & 32.16          & 32.12          \\ \hline
PAN            & 261             & 160              & 38.00          & 33.59          & 32.18          & 32.01          \\ \hline
FENet          & 675             & -                & 38.08          & 33.70          & 32.20          & 32.18          \\ \hline
MRDN           & 565             & 288              & 38.05          & 33.68          & 32.24          & 32.42          \\ \hline
LatticeNet     & 756             & 381              & 38.15          & 33.78          & 32.25          & 32.43          \\ \hline
ADBNet         & 535             & 225              & 38.25          & 34.10          & 32.39          & 33.31          \\ \hline
\end{tabular}

$^a$ This model has it own hardware design. \\
$^*$ This model has only $\times$3 and we retrained it for $\times$2.

\end{minipage}
\hfill
\begin{minipage}{\columnwidth}
\centering
\caption {The quantitative results of several light weight methods($\times$4). The bold one represents our method.}
\label{Comparison of several light weight models. x4}
\begin{tabular}{|c|c|c|c|c|c|c|c|}
\hline
\textbf{Model} & \textbf{Size(K)} & \textbf{GMAC} & \textbf{Set5}  & \textbf{Set14} & \textbf{B100}  & \textbf{U.100}      \\ \hline
Bicubic        & -               & -                & 28.42          & 26.00          & 25.96          & 23.14          \\ \hline
FSRCNN-s       & 4               & 6                & 30.11          & 27.19          & 26.84          & -              \\ \hline
$^a$SRNPU          & 183             & 1                & 30.41          & 27.37          & 26.86          & -          \\ \hline
SRCNN          & 57              & 119              & 30.48          & 27.49          & 26.90          & 24.52          \\ \hline
FSRCNN         & 13              & 10               & 30.71          & 27.59          & 26.98          & 24.62          \\ \hline
XLSR*          & 28              & 4                & 30.69          & 27.70          & 26.98          & 24.58          \\ \hline
\textbf{ACNet} & \textbf{18}     & \textbf{2}       & \textbf{30.78} & \textbf{27.62} & \textbf{27.00} & \textbf{24.63} \\ \hline
HPAN           & 26              & 8                & 30.88          & 27.68          & 27.03          & 24.69          \\ \hline
VDSR           & 665             & 1,225            & 31.35          & 28.01          & 27.29          & 25.18          \\ \hline
LapSRN         & 813             & 336              & 31.54          & 28.19          & 27.32          & 25.21          \\ \hline
LWN            & 286             & 39               & 31.70          & 28.19          & 27.37          & 25.48          \\ \hline
DRRN           & 297             & 15,293           & 31.68          & 28.21          & 27.38          & 25.44          \\ \hline
IDN            & 600             & 72               & 31.82          & 28.25          & 27.41          & 25.41          \\ \hline
CARN-M         & 412             & 73               & 31.92          & 28.42          & 27.44          & 25.62          \\ \hline
MADNet         & 1,002           & 122              & 32.01          & 28.45          & 27.47          & 25.77          \\ \hline
T-FMB          & 690             & -                & 32.08          & 28.51          & 27.49          & 25.89          \\ \hline
RFDN           & 643             & 72               & 32.24          & 28.61          & 27.57          & 26.11          \\ \hline
PAN            & 272             & 64               & 32.13          & 28.61          & 27.59          & 26.11          \\ \hline
$^a$eCNN           & 583             & 398              & 32.11          & 28.61          & 27.59          & 26.11      \\ \hline
LatticeNet     & 777             & 98               & 32.30          & 28.68          & 27.62          & 26.25          \\ \hline
FENet          & 675             & -                & 32.24          & 28.61          & 27.63          & 26.20          \\ \hline
MRDN           & 582             & 72               & 32.36          & 28.76          & 27.67          & 26.41          \\ \hline
ADBNet         & 535             & 57               & 32.82          & 29.01          & 27.95          & 27.16          \\ \hline
\end{tabular}

$^a$ This model has its own hardware design. 
\end{minipage}
\end{table*}

\subsection{Ablation Study}
\subsubsection{Model depth}
\begin{figure}[htbp]
    \centering
    \includegraphics[height=!,width=1.0\linewidth,keepaspectratio=true]{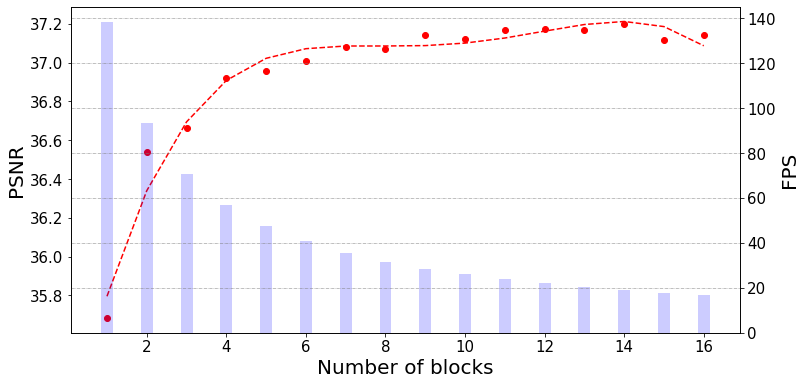}
    \caption {The trend line of increasing block numbers and the corresponding frame rate in scaling factor $\times$2.}
    \label{The trend line of increasing block numbers}
\end{figure}

A direct way to increase the depth of our model is to stack more CBBs. We explore different blocks and show the results in Fig.~\ref{The trend line of increasing block numbers}. We trained every model for only 600 epochs, and the other settings are the same as in the previous setup. As the depth deepens, the performance becomes saturated around 37.2 dB. However, when we stack 60 blocks in the model, because of the gradient vanishing, the model cannot learn anything. If we want to improve it, we have to add some skip connections in the model.

\subsubsection{Decoupled asymmetric convolution}
Unlike \cite{8429522} which uses both asymmetric and normal convolutions, our model uses only decoupled asymmetric convolutions and 1x1 convolutions. 
Table~\ref{The results of different configurations of asymmetric convolution} shows the different configurations of the asymmetric convolution.
As shown in Table~\ref{The results of different configurations of asymmetric convolution}, the first two rows are models with only one type of asymmetric convolution, and the last two rows are models with one type of convolution in the first stage and the other type in the rest of the stages. Because such models lack a receptive field in another direction, the recovered images are worse than the last two rows. The results of the last two rows show that simply using one layer of different types of convolution can improve the performance of the reconstruction capacity.

\begin{table}[htbp]
\centering
\caption{Results of different configurations of asymmetric convolution.}
\label{The results of different configurations of asymmetric convolution}
\begin{tabular}{|c|ccc|}
\hline
                                   & Set5  & B100  & M.109 \\ \hline
All   vertical                     & 34.55 & 30.00 & 32.07 \\
All   horizontal                   & 34.06 & 30.10 & 31.57 \\ \hline
First   horizontal , rest vertical & 37.21 & 31.61 & 36.59 \\
First   vertical , rest horizontal & 37.19 & 31.59 & 36.66 \\ \hline
\end{tabular}

\end{table}

\subsection{Hardware}
\subsubsection{Implementation results and design comparison}
Table~\ref{Comparison with other designs table SR} shows our design implementation and comparison to other works. Our design in the TSMC 40nm CMOS process can attain real-time full HD SR with 2332K gate count and 198KB SRAM when operating at 270MHz. Our design needs the smallest external bandwidth due to the \textit{holistic model fusion}, which is enabled by our small model size.  

Compared to other designs, the proposed design has low MAC numbers and small buffer size, but better performance due to the deeper structure, especially compared to other FSRCNN-based designs\cite{9159619, 9401206}. Our energy efficiency is also better than others except \cite{huang2019ecnn}. PSNR in~\cite{huang2019ecnn} is 0.5 dB higher due to their 583K-parameter model. Due to this large model, its throughput becomes lower. Its energy efficiency is higher than that of ours because of its large number of MACs. SRNPU~\cite{9159619} has a similar throughput as ours in the scaling factors $\times$2, but the throughput in the scaling factor $\times$4 is lower than ours. The main reason is that they need to support different models in one architecture. In its design, the PE utilization is 88.2 \% for the large model but it drops to only 51 \% for the simple model. In contrast, as shown in Fig.\ref{measured in x2 scaling factor}, our average PE utilization in scaling factors $\times$2 and $\times$4 is almost the same. Our energy efficiency is at least 1.8x higher than that of the SRNPU.

\begin{table*}[htbp]
\caption{Comparison with other designs, where "-" denotes data unavailable.}
\label{Comparison with other designs table SR}
\begin{tabular}{|cccccc|}
\hline
\multicolumn{1}{|c|}{}                                                                       & \multicolumn{1}{c|}{ACNPU}                                                                               & \multicolumn{1}{c|}{eCNN~\cite{huang2019ecnn}}                                                                                         & \multicolumn{1}{c|}{SRNPU~\cite{9159619}}                                                                                               & \multicolumn{1}{c|}{~\cite{9073972}}                                                                  & FXPU~\cite{9401206}                                                                 \\ \hline
\multicolumn{1}{|c|}{Process}                                                                & \multicolumn{1}{c|}{40 nm}                                                                               & \multicolumn{1}{c|}{40 nm}                                                                                        & \multicolumn{1}{c|}{65 nm}                                                                                               & \multicolumn{1}{c|}{32 nm}                                                              & 28 nm                                                                \\ \hline
\multicolumn{1}{|c|}{Method}                                                                 & \multicolumn{1}{c|}{ACNet}                                                                       & \multicolumn{1}{c|}{ERNet}                                                                                        & \multicolumn{1}{c|}{\begin{tabular}[c]{@{}c@{}}Tile-based Selective \\ FSRCNN\end{tabular}}                              & \multicolumn{1}{c|}{Modified IDN}                                                       & FSRCNN                                                               \\ \hline
\multicolumn{1}{|c|}{Precision(W/A)(bits)}                                                   & \multicolumn{1}{c|}{FP10 / FP13}                                                                         & \multicolumn{1}{c|}{FXP8 / FXP8}                                                                                  & \multicolumn{1}{c|}{FXP8 / FXP16}                                                                                        & \multicolumn{1}{c|}{- / FXP12}                                                          & FXP8+FP16                                                            \\ \hline
\multicolumn{1}{|c|}{$^a$PSNR(dB)}                                                           & \multicolumn{1}{c|}{37.30}                                                                               & \multicolumn{1}{c|}{37.80}                                                                                        & \multicolumn{1}{c|}{37.06}                                                                                               & \multicolumn{1}{c|}{36.96}                                                              & -                                                                    \\ \hline
\multicolumn{1}{|c|}{Parameter   (K)}                                                        & \multicolumn{1}{c|}{\textbf{18}}                                                                                  & \multicolumn{1}{c|}{583}                                                                                          & \multicolumn{1}{c|}{183}                                                                                                 & \multicolumn{1}{c|}{13}                                                                 & 13                                                                   \\ \hline
\multicolumn{1}{|c|}{Supported   Scale}                                                      & \multicolumn{1}{c|}{×2, ×4}                                                                              & \multicolumn{1}{c|}{×2, ×4}                                                                                       & \multicolumn{1}{c|}{×2, ×3, ×4}                                                                                          & \multicolumn{1}{c|}{×2}                                                                 & -                                                                    \\ \hline
\multicolumn{1}{|c|}{Layer   Fusion}                                                         & \multicolumn{1}{c|}{\begin{tabular}[c]{@{}c@{}}Holistic \\ Model Fusion\end{tabular}}               & \multicolumn{1}{c|}{\begin{tabular}[c]{@{}c@{}}Overlap Tiling \\ Layer Fusion\end{tabular}}                       & \multicolumn{1}{c|}{\begin{tabular}[c]{@{}c@{}}Selective Caching-based \\ Layer   Fusion\end{tabular}}                   & \multicolumn{1}{c|}{Layer Fision}                                                       & -                                                                    \\ \hline
\multicolumn{1}{|c|}{External   Bandwidth}                                                   & \multicolumn{1}{c|}{\textbf{I/O}}                                                                                 & \multicolumn{1}{c|}{I/O + Intermediate}                                                                           & \multicolumn{1}{c|}{I/O + Intermediate}                                                                                  & \multicolumn{1}{c|}{-}                                                                  & -                                                                    \\ \hline
\multicolumn{1}{|c|}{Frequency(MHz)}                                                         & \multicolumn{1}{c|}{270}                                                                           & \multicolumn{1}{c|}{250}                                                                                          & \multicolumn{1}{c|}{$\sim$200}                                                                                           & \multicolumn{1}{c|}{200}                                                                & 250                                                                  \\ \hline
\multicolumn{1}{|c|}{Gate   Count(K)}                                                        & \multicolumn{1}{c|}{\textbf{2332}}                                                                                & \multicolumn{1}{c|}{-}                                                                                            & \multicolumn{1}{c|}{-}                                                                                                   & \multicolumn{1}{c|}{3114}                                                               & -                                                                    \\ \hline
\multicolumn{1}{|c|}{SRAM(KB)}                                                               & \multicolumn{1}{c|}{\textbf{198}}                                                                                 & \multicolumn{1}{c|}{2,864}                                                                                        & \multicolumn{1}{c|}{572}                                                                                                 & \multicolumn{1}{c|}{-}                                                                  & 215                                                                  \\ \hline
\multicolumn{1}{|c|}{MAC   Number}                                                           & \multicolumn{1}{c|}{1,248}                                                                               & \multicolumn{1}{c|}{81,920}                                                                                       & \multicolumn{1}{c|}{1,152}                                                                                               & \multicolumn{1}{c|}{1,152}                                                              & -                                                                    \\ \hline
\multicolumn{1}{|c|}{SR Speed   (fps)}                                                       & \multicolumn{1}{c|}{\begin{tabular}[c]{@{}c@{}}31.7(×2, Full HD)\\      124.4(×4, Full HD)\end{tabular}} & \multicolumn{1}{c|}{\begin{tabular}[c]{@{}c@{}}60.0/$^c$26.6(×2,   HD)\\      30.0/$^c$13.3(×4, HD)\end{tabular}} & \multicolumn{1}{c|}{\begin{tabular}[c]{@{}c@{}}31.8(×2, Full HD)\\       88.3(×4, Full HD)\end{tabular}}                 & \multicolumn{1}{c|}{\begin{tabular}[c]{@{}c@{}}60.0(×2, Full HD)\\      -\end{tabular}} & \begin{tabular}[c]{@{}c@{}}-\\      74.6(×4, Full HD)\end{tabular}   \\ \hline
\multicolumn{1}{|c|}{Power (W)}                                                              & \multicolumn{1}{c|}{\begin{tabular}[c]{@{}c@{}}0.1318\\      (@0.9V,270MHz)\end{tabular}}                & \multicolumn{1}{c|}{\begin{tabular}[c]{@{}c@{}}6.95$\sim$7.24\\      (@0.9V,250MHz)\end{tabular}}                 & \multicolumn{1}{c|}{\begin{tabular}[c]{@{}c@{}}0.2110\\      (@1.1V,200MHz)\end{tabular}}                                & \multicolumn{1}{c|}{-}                                                                  & \begin{tabular}[c]{@{}c@{}}0.1333\\      (@0.9V,250MHz)\end{tabular} \\ \hline
\multicolumn{1}{|c|}{\begin{tabular}[c]{@{}c@{}}Energy   Efficiency\\ (TOPs/W)\end{tabular}} & \multicolumn{1}{c|}{4.75/$^b$4.75}                                                                       & \multicolumn{1}{c|}{\begin{tabular}[c]{@{}c@{}}5.49/$^b$5.49(×2)\\      5.87/$^b$5.87(×4)\end{tabular}}           & \multicolumn{1}{c|}{\begin{tabular}[c]{@{}c@{}}1.9/$^b$2.14  (@0.75V,50MHz)\\ 1.1/$^b$2.67  (@1.1V,200MHz)\end{tabular}} & \multicolumn{1}{c|}{-}                                                                  & 3.6/$^b$2.52                                                         \\ \hline
\multicolumn{6}{|l|}{\begin{tabular}[c]{@{}l@{}}$^a$Performance on Set5\\ $^b$Normalized energy efficiency = energy efficiency $\times \left ( \frac{process}{40nm} \right )\times \left ( \frac{voltage}{0.9V} \right )^{2}$.\\ $^c$Normalizing HD FPS to FHD FPS\\ I/O means the bandwidth contains only the input image and the output image.\end{tabular}}                                                                                                                                                                                                                                                          \\ \hline
\end{tabular}

\end{table*}

\subsubsection{Hardware utilization}

\begin{figure}[htbp]
 \subfloat[Execution time ]{\includegraphics[trim={2.5cm 0 2.5cm 0}, height=!,width=0.43\linewidth,keepaspectratio=true]{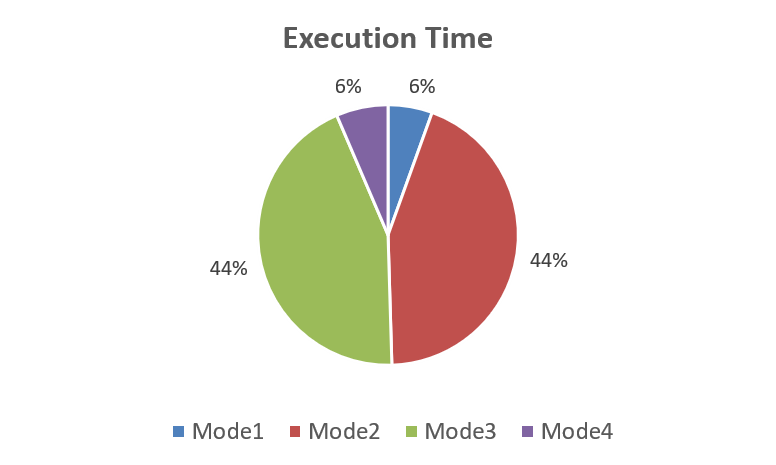}}
 \hspace{0.5cm}
 \subfloat[Hardware utilization]{\includegraphics[trim={2cm 0 2cm 0}, height=!,width=0.43\linewidth,keepaspectratio=true]{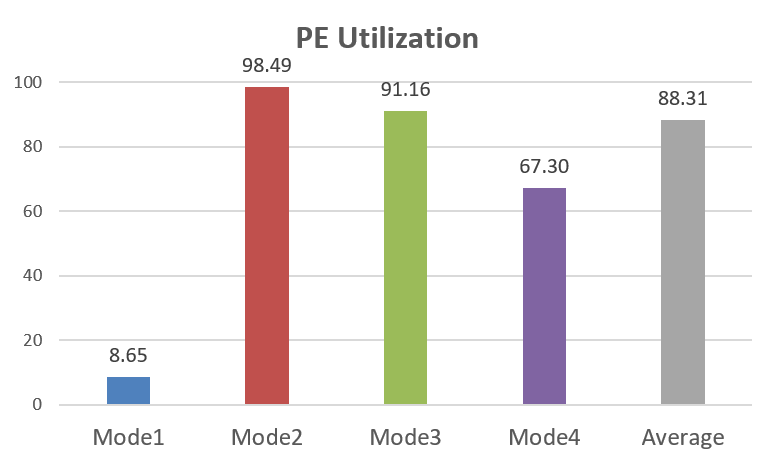}}

\caption {The utilization breakdown for different modes, measured in $\times$2 scaling factor.}
\label{measured in x2 scaling factor}
\end{figure}

Fig.~\ref{measured in x2 scaling factor} shows the execution time and hardware utilization of four modes for $\times$2 scaling. A similar distribution is also found in $\times$4 scaling. As mentioned before, although the utilization of the first mode is very low, its execution time is insignificant for the whole model. Therefore, the average PE utilization is about 88.3 \% and 88.0 \% at $\times$2 and $\times$4, respectively. The reason why the utilization at $\times$4 is lower than at $\times$2 is that the execution time of mode 4 is 2 \% more and thus decreases the average utilization.

\subsubsection{Area analysis}
\begin{figure}[htbp]
 \subfloat[Area breakdown]{\includegraphics[trim={3cm 0 3cm 0}, height=!,width=0.4\linewidth,keepaspectratio=true]{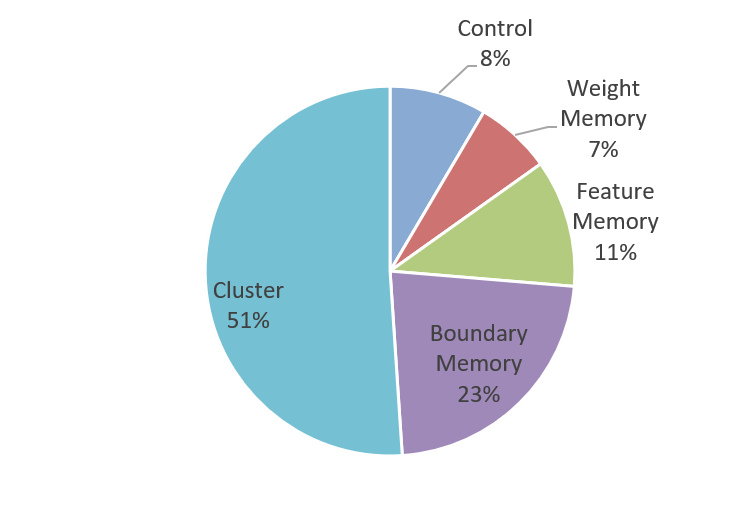}}
 \hspace{1cm}
 \subfloat[Power breakdown]{\includegraphics[trim={2.5cm 0 2.5cm 0}, height=!,width=0.31\linewidth,keepaspectratio=true]{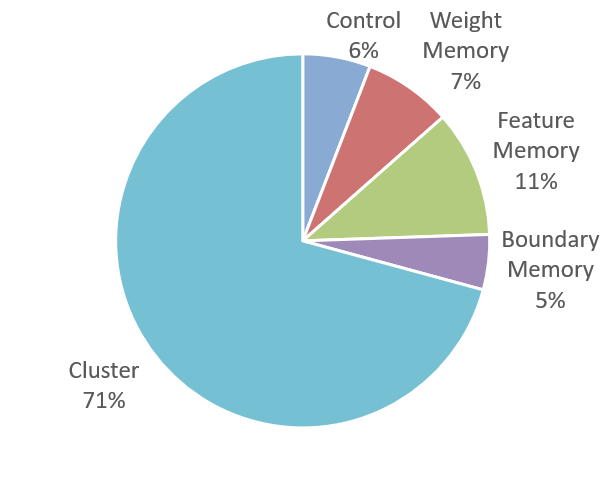}}

    \caption {The area and power breakdown of ACNPU.}
    \label{The area breakdown of ACNPU}
\end{figure}

Fig.~\ref{The area breakdown of ACNPU} shows the area analysis of our design. Because our ACNet is an ultra-light-weight model, the area of the weight memory is small. The area of the feature memory is smaller than the area of the boundary memory because we have to handle all partial boundary sums.  The size of the boundary memory will be about $\times$64 larger if we do not use the asymmetric architecture. The cluster area is the largest because there are many floating-point multiplication and addition units with buffers in all clusters.

\subsubsection{Power analysis}
Despite having the second largest size, the boundary memory only uses 5\% of the total power because it will be accessed only in three 3×1 convolution layers. The power of the feature memory is only 11\% due to the stationary of the input features. The six clusters work at every stage and, as a result, consume most of the power. 

\subsubsection{Memory bandwidth}
As mentioned above, the proposed architecture fuses the whole model for on-chip execution. Compared to a layer-by-layer accelerator, ACNPU can completely eliminate the additional bandwidth of off-chip feature access shown in Table~\ref{Comparison with the layer-by-layer}. The values are measured in $\times$2 Full HD 30 FPS. Furthermore, we execute 1$\times$1 and 1$\times$3 in parallel to reduce access to on-chip feature memory, which reduces the amount of on-chip memory access by 36.4 \%.

\begin{table}[htbp]
\centering
\caption{Comparison with the layer-by-layer processing and our \textit{holistic model fusion}.}
\label{Comparison with the layer-by-layer}
\begin{tabular}{|c|c|c|}
\hline
Off-chip   bandwidth & Layer-by-layer & Holistic model fusion \\ \hline
Bandwidth   (GB/s)   & 35.709         & 0.126                        \\ \hline
\end{tabular}

\end{table}

\label{chapter:conclusion}

\section{Conclusion}

This paper realizes an energy-efficient SR accelerator, ACNPU, for resource-limited edge devices to achieve low power consumption and high image quality. Compared to FSRCNN, the ACNPU uses a 27-layer model for 0.34dB higher quality, but costs 36\% less complexity with similar model size due to \textit{decoupled asymmetric convolution and split-bypass structure}. The model structure is also hardware-friendly, as it employs the aforementioned minimal structures and local connections rather than long connections. With this hardware-friendly structure and 17K model size, the accelerator can eliminate external memory access of intermediate feature maps by the \textit{holistic model fusion}. Internal memory access is further reduced by parallel execution of 1x3/1x1 layers and \textit{local input stationary flow}. The corresponding hardware is regular and easy to control to support different layers by \textit{PE clusters with adaptive input values and uniform data flow}. The final implementation with the TSMC 40nm CMOS process can achieve full HD processing with 31.7 and 124.4 frames per second for scaling $\times$2 and $\times$4 at the working frequency of 270 MHz, respectively, achieving 4.75 TOPs/W energy efficiency and being 1.8$times$ higher than the previous SR accelerator.

\bibliographystyle{IEEEtran}

\bibliography{IEEEabrv,bib/thesis}

\begin{thebibliography}{10}
\providecommand{\url}[1]{#1}
\csname url@samestyle\endcsname
\providecommand{\newblock}{\relax}
\providecommand{\bibinfo}[2]{#2}
\providecommand{\BIBentrySTDinterwordspacing}{\spaceskip=0pt\relax}
\providecommand{\BIBentryALTinterwordstretchfactor}{4}
\providecommand{\BIBentryALTinterwordspacing}{\spaceskip=\fontdimen2\font plus
\BIBentryALTinterwordstretchfactor\fontdimen3\font minus
  \fontdimen4\font\relax}
\providecommand{\BIBforeignlanguage}[2]{{%
\expandafter\ifx\csname l@#1\endcsname\relax
\typeout{** WARNING: IEEEtran.bst: No hyphenation pattern has been}%
\typeout{** loaded for the language `#1'. Using the pattern for}%
\typeout{** the default language instead.}%
\else
\language=\csname l@#1\endcsname
\fi
#2}}
\providecommand{\BIBdecl}{\relax}
\BIBdecl

\bibitem{dong2015image}
C.~Dong \emph{et~al.}, ``Image super-resolution using deep convolutional
  networks,'' \emph{IEEE {T}ransactions on {P}attern {A}nalysis and {M}achine
  {I}ntelligence}, vol.~38, no.~2, pp. 295--307, 2015.

\bibitem{lai2017deep}
W.-S. Lai \emph{et~al.}, ``Deep laplacian pyramid networks for fast and
  accurate super-resolution,'' in \emph{Proceedings of the IEEE {C}onference on
  {C}omputer {V}ision and {P}attern {R}ecognition}, 2017, pp. 624--632.

\bibitem{zhang2018image}
Y.~Zhang \emph{et~al.}, ``Image super-resolution using very deep residual
  channel attention networks,'' in \emph{Proceedings of the European
  {C}onference on {C}omputer {V}ision (ECCV)}, 2018, pp. 286--301.

\bibitem{liu2020residual}
J.~Liu \emph{et~al.}, ``Residual feature aggregation network for image
  super-resolution,'' in \emph{Proceedings of the IEEE/CVF {C}onference on
  {C}omputer {V}ision and {P}attern {R}ecognition}, 2020, pp. 2359--2368.

\bibitem{9223656}
S.~Lee \emph{et~al.}, ``{CNN} acceleration with hardware-efficient dataflow for
  super-resolution,'' \emph{IEEE Access}, vol.~8, pp. 187\,754--187\,765, 2020.

\bibitem{9073972}
P.-W. Yen \emph{et~al.}, ``Real-time super resolution {CNN} accelerator with
  constant kernel size winograd convolution,'' \emph{2nd IEEE International
  Conference on Artificial Intelligence Circuits and Systems (AICAS)}, pp.
  193--197, 2020.

\bibitem{9159619}
J.~Lee, J.~Lee, and H.-J. Yoo, ``{SRNPU}: An energy-efficient {CNN}-based
  super-resolution processor with tile-based selective super-resolution in
  mobile devices,'' \emph{IEEE Journal on Emerging and Selected Topics in
  Circuits and Systems}, vol.~10, no.~3, pp. 320--334, 2020.

\bibitem{huang2019ecnn}
C.-T. Huang \emph{et~al.}, ``e{CNN}: A block-based and highly-parallel {CNN}
  accelerator for edge inference,'' in \emph{Proceedings of the 52nd Annual
  IEEE/ACM International Symposium on Microarchitecture}, 2019, pp. 182--195.

\bibitem{9401206}
Z.~Li \emph{et~al.}, ``A 3.6 {TOPS/W} hybrid {FP}-{FXP} deep learning processor
  with outlier compensation for image-to-image application,'' in \emph{IEEE
  International Symposium on Circuits and Systems (ISCAS)}, 2021, pp. 1--5.

\bibitem{BSRA}
\BIBentryALTinterwordspacing
D.-H. Yang and T.-S. Chang, ``{BSRA}: Block-based super resolution accelerator
  with hardware efficient pixel attention,'' in \emph{IEEE International
  Symposium on Circuits and Systems (ISCAS)}, 2022. [Online]. Available:
  \url{https://arxiv.org/abs/2205.00777}
\BIBentrySTDinterwordspacing

\bibitem{dong2016accelerating}
C.~Dong, C.~C. Loy, and X.~Tang, ``Accelerating the super-resolution
  convolutional neural network,'' in \emph{European {C}onference on {C}omputer
  {V}ision}.\hskip 1em plus 0.5em minus 0.4em\relax Springer, 2016, pp.
  391--407.

\bibitem{8429522}
Y.~Kim, J.-S. Choi, and M.~Kim, ``A real-time convolutional neural network for
  super-resolution on {FPGA} with applications to 4{K} {UHD} 60 {FPS} video
  services,'' \emph{IEEE Transactions on Circuits and Systems for Video
  Technology}, vol.~29, no.~8, pp. 2521--2534, 2019.

\bibitem{DBLP:journals/corr/abs-2105-10288}
\BIBentryALTinterwordspacing
M.~Ayazoglu, ``Extremely lightweight quantization robust real-time single-image
  super resolution for mobile devices,'' \emph{CoRR}, vol. abs/2105.10288,
  2021. [Online]. Available: \url{https://arxiv.org/abs/2105.10288}
\BIBentrySTDinterwordspacing

\bibitem{ding2019acnet}
X.~Ding \emph{et~al.}, ``Acnet: Strengthening the kernel skeletons for powerful
  {CNN} via asymmetric convolution blocks,'' in \emph{Proceedings of the
  IEEE/CVF international conference on computer vision}, 2019, pp. 1911--1920.

\bibitem{tian2021asymmetric}
C.~Tian \emph{et~al.}, ``Asymmetric {CNN} for image superresolution,''
  \emph{IEEE Transactions on Systems, Man, and Cybernetics: Systems}, vol.~52,
  no.~6, pp. 3718--3730, 2021.

\bibitem{7783725}
M.~Alwani \emph{et~al.}, ``Fused-layer {CNN} accelerators,'' in \emph{49th
  Annual IEEE/ACM International Symposium on Microarchitecture (MICRO)}, 2016,
  pp. 1--12.

\bibitem{bevilacqua2012low}
M.~Bevilacqua \emph{et~al.}, ``Low-complexity single-image super-resolution
  based on nonnegative neighbor embedding,'' \emph{{B}ritish {M}achine {V}ision
  {C}onference, {S}urrey, {UK}, 3-7 September}, 2012.

\bibitem{yang2010image}
J.~Yang \emph{et~al.}, ``Image super-resolution via sparse representation,''
  \emph{IEEE {T}ransactions on {I}mage {P}rocessing}, vol.~19, no.~11, pp.
  2861--2873, 2010.

\bibitem{martin2001database}
D.~Martin \emph{et~al.}, ``A database of human segmented natural images and its
  application to evaluating segmentation algorithms and measuring ecological
  statistics,'' in \emph{Proceedings Eighth IEEE International Conference on
  Computer Vision. ICCV}, vol.~2, 2001, pp. 416--423.

\bibitem{huang2015single}
J.-B. Huang, A.~Singh, and N.~Ahuja, ``Single image super-resolution from
  transformed self-exemplars,'' in \emph{Proceedings of the IEEE {C}onference
  on {C}omputer {V}ision and {P}attern {R}ecognition}, 2015, pp. 5197--5206.

\bibitem{LWN}
J.~Y. Park, D.~Y. Choi, and B.~C. Song, ``Slice-based super-resolution using
  light-weight network with relation loss,'' in \emph{IEEE International
  Conference on Image Processing (ICIP)}, 2020, pp. 503--507.

\end{thebibliography}

\begin{IEEEbiography}[{\includegraphics[width=1in,height=1.25in,clip,keepaspectratio]{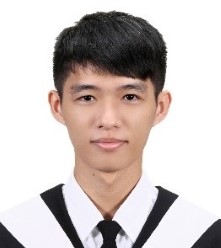}}]{Tun-Hao Yang}
received the M.S. degree in electronics engineering from the National Yang Ming Chiao Tung University, Hsinchu, Taiwan, in 2022. He is currently working in the Richtek, Hsinchu, Taiwan. His research interest includes superresolution neural network and VLSI design.

\end{IEEEbiography}

\begin{IEEEbiography}[{\includegraphics[width=1in,height=1.25in,clip,keepaspectratio]{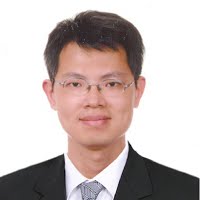}}]{Tian-Sheuan Chang}
	(S’93–M’06–SM’07)
	received the B.S., M.S., and Ph.D. degrees in electronic engineering from National Chiao-Tung University (NCTU), Hsinchu, Taiwan, in 1993, 1995, and 1999, respectively. 
	
	From 2000 to 2004, he was a Deputy Manager with Global Unichip Corporation, Hsinchu, Taiwan. In 2004, he joined the Department of Electronics Engineering, NCTU (as National Yang Ming Chiao Tung University (NYCU) in 2021), where he is currently a Professor. In 2009, he was a visiting scholar in IMEC, Belgium. His current research interests include system-on-a-chip design, VLSI signal processing, and computer architecture.
	
	Dr. Chang has received the Excellent Young Electrical Engineer from Chinese Institute of Electrical Engineering in 2007, and the Outstanding Young Scholar from Taiwan IC Design Society in 2010. He has been actively involved in many international conferences as an organizing committee or technical program committee member.
\end{IEEEbiography}
\end{document}